\documentclass{elsarticle}
\usepackage{graphicx,color}
\usepackage{amsmath}

\begin{document}
\begin{frontmatter}
\title{The Extraordinary Glow}
\author[ind,rbrc]{Jinfeng Liao \fnref{fn}}

\fntext[fn]{liaoji@indiana.edu}

\address[ind]{Physics Department and Center for Exploration of Energy and Matter,
Indiana University, \\2401 N Milo B. Sampson Lane, Bloomington, IN 47408, USA.}
\address[rbrc]{RIKEN BNL Research Center, Bldg. 510A, Brookhaven National Laboratory,\\
   Upton, NY 11973, USA.}

\begin{abstract}
In this contribution I discuss some recent progress in understanding the evolution of the pre-thermal quark-gluon matter, known as the glasma, during the early stage in heavy ion collisions, and the implication for early time photon and dilepton  emissions. 
\end{abstract}
\end{frontmatter}

\section{Introduction}

This is a contribution to the memorial volume as proceedings for the conference ``45 Years of Nuclear Theory at Stony Brook: A Tribute to Gerald E. Brown''. I came to study for PhD at Stony Brook in 2004 and two years later officially became a student   in the Nuclear Theory Group. Unlike many other contributors in this volume, I was neither a student/postdoc nor a collaborator of Gerry --- at the time Edward Shuryak was my advisor. But like all the other contributors, I benefited so much from Gerry through the numerous physics discussions and personal interactions.  The great care, the generous help, and the extreme kindness that he bestowed upon young people like me,  was ``the extraordinary glow'' that fostered the growth of generations of nuclear theorists, and that I was really lucky to enjoy and appreciate.

The last physics discussions I had with Gerry were about the photon and dilepton measurements from PHENIX Collaboration at Relativistic Heavy Ion Collider (RHIC). That was during my last few months of PhD and I was about to leave for my first postdoc position at LBNL, for which  Gerry gave me a great deal of help in my application. In short, the PHENIX measurements show large excess of low-to-intermediate mass dileptons in AuAu collisions, beyond contributions from all usual emission sources considered at the time~\cite{Adare:2008fqa,Adare:2009qk,Adare:2011zr}. Gerry had always been very interested in the behavior of meson masses in the temperature regime $120\sim 170 \, \rm Mev$ particularly related to chiral symmetry restoration. He had the idea that such excessive dileptons may come from a ``sticky-pion'' picture, in which very light pions merge in specific way into similarly light rho mesons which then decay into soft dileptons. We spent quite many hours discussing such processes and he encouraged me to talk with Volker Koch further when I arrived at LBNL and to work together on this problem. That did not go very far, particularly after the very unfortunate health situation of Gerry that occured just few months after my graduation. A few years later, I came back to this problem about ``the extraordinary glow'' in heavy ion collisions,  from a quite different angle~\cite{Chiu:2012ij}. While the idea of Gerry pertains to the emission from the very late stage ( i.e. the hadronic gas ) of evolution, the more recent development that I will discuss in the rest of this article has the emission origin from the very early stage (i.e. the pre-thermal glasma) of evolution. Gerry and I could have enjoyed a  delightful discussion  on this if he were still with us.

\section{The overpopulated glasma}

Thermalization of the quark-gluon plasma is one of the most
challenging problems in current heavy ion physics~\cite{Berges:2012ks,Huang:2014iwa,Gelis:2012ri,Fukushima:2011ca,Strickland:2013uga}.  Starting with  two
colliding nuclei in a form of color glass condensate, the system shortly after the initial 
impact becomes a very dense, far-from-equilibrium system of gluons called the ``glasma''. 
The glasma has high gluon
occupation $1/\alpha_{\rm s}$ up to the saturation scale $Q_{\rm s}$ of few GeV (implying a viable 
weak coupling approach). Extensive phenomenological studies 
suggest that the evolution of this glasma stage toward a hydrodynamically expanding 
quark-gluon plasma occurs  at a time of the order of $1$ fermi/c. Exactly how this happens remains a big puzzle.

Since the initial scale $Q_s$ in the glasma is large and thus the coupling is weak, the kinetic theory seems to be a natural and plausible framework to investigate the detailed evolution of the phase space distribution in the dense gluon system starting from the time scale $\sim 1/Q_{\rm s}$. Such efforts were initiated long ago~\cite{Mueller:1999fp,Mueller:1999pi,Baier:2000sb,Baier:2002qv,Arnold:2002zm,Mueller:2005un,Mueller:2005hj,Mueller:2006up,Xu:2004mz}, with many highly interesting developments in the past few years:  see e.g. review and references in \cite{Berges:2012ks,Huang:2014iwa}. An apparent tension in such approaches exists in that in a naive counting the scattering rate (of leading elastic processes) $\sim \alpha_s^2$ may not be able to bring the system back to thermalization quickly enough. A number of past kinetic works suggest that the inelastic processes may play more significant role as compared with the elastic ones in speeding up the thermalization process, especially in populating the very soft momentum region. This may be true in the dilute regime (close to the Boltzmann limit), however may not be the accurate picture when the system under consideration is in the {\it highly overpopulated regime} with $f\sim 1/\alpha_s$.  As shown in a number of recent kinetic studies~\cite{Blaizot:2011xf,Blaizot:2013lga,Huang:2013lia,Blaizot:2014jna,BJL,Ruggieri:2013ova}, the elastic scatterings with highly overpopulated initial conditions can lead to order $\sim {\alpha_s^0}$ evolution and develop strong infrared cascade with the Bose enhancement, and in fact may even induce a dynamical Bose-Einstein Condensation. (In passing let us mention that there have also been lots of developments  in understanding the evolution of overpopulated glasma with different approaches, see e.g. \cite{Gelis:2013rba,Berges:2012us,Berges:2013fga,Kurkela:2011ub}.) In the present Section let us briefly discuss some of these nontrivial features of the glasma as a basis for our discussions of its electromagnetic emissions in the next Section.

\subsection{The highly overpopulated Glasma}

To see a few nontrivial features associated with high overpopulation, let us consider the kinetic evolution in a weakly coupled gluon system  initially described by the following glasma-type distribution as inspired by the CGC picture:
\begin{eqnarray} \label{eq_f0}
f(p \leq Q_{\rm s}) = f_0 \quad , \quad f(p>Q_{\rm s}) = 0 \, .
\end{eqnarray}
For the glasma in heavy ion collisions, the phase space is maximally filled: $f_0\sim 1/\alpha_s$ (with $\alpha_{\rm s}\ll 1$). As pointed out in~\cite{Blaizot:2011xf}, such high occupation coherently amplifies scattering and changes usual power counting of scattering rate:
 the resulting collision term from the $2\leftrightarrow 2$ gluon scattering process will scale as $\sim
\alpha_{\rm s}^2 f^2 \sim \hat{o}(1)$  despite smallish $\alpha_{\rm s}$.  This is a natural consequence of the essential Bose enhancement factor $(1+f)$ which would scale as $f$ in the dense regime while scale as $1$ in the dilute regime. With the coupling constant dropping out of the problem,    the system
  behaves as an emergent strongly interacting matter, even though the elementary coupling is small.  
  
  This point may be elaborated also in another way by examining a plasma coupling parameter $\Gamma$ defined as the ratio between potential energy density $P$ due to interaction and the kinetic energy density $K$ from thermal motion: $\Gamma=P/K$. For a weakly coupled system of gluons with typical occupation $f$ and typical scale $Q$, the kinetic energy density scales as $K\sim f Q^4$ while the interaction energy density scales as $P\sim f^2 Q^3 \alpha_{\rm s} / (1/Q)\sim f^2\alpha_{\rm s} Q^4$. Thus the ratio scales as $\Gamma \sim f \alpha_{\rm s}$. While normally (e.g. in thermal plasma) $f\sim 1$ and $\Gamma\sim \alpha_{\rm s} <<1$, the glasma with high occupation $f\sim 1/\alpha_{\rm s}$ will have a plasma coupling $\Gamma\sim 1$ despite the smallish coupling. Extensive studies over various many-body systems in the past have shown that the gas regime occurs with $\Gamma <<1$ while a liquid regime emerges universally at $\Gamma \sim 1$ (see e.g. \cite{Gelman:2006xw,Liao:2006ry}). This therefore again hints at a strongly interacting glasma that may exhibit fluid-like behavior by virtue of high population.

A novel finding ~\cite{Blaizot:2011xf,Blaizot:2013lga}, hitherto unrealized, is that a system with such initial condition is highly overpopulated: that is, the gluon occupation number is parametrically large  when compared to a system in thermal equilibrium with the same
  energy density.   To illustrate this point, consider the energy and particle number densities with the initial distribution (\ref{eq_f0}), we have
\begin{eqnarray}\label{initialparam}
 \epsilon_0  = f_0\, \frac{Q_s^4}{  8\pi^2},\qquad n_0=f_0\, \frac{Q_s^3}{6\pi^2},\qquad
n_0 \, \epsilon_0^{-3/4} = f_0^{1/4} \, \frac{2^{5/4}}{3\, \pi^{1/2}} ,
\end{eqnarray}
with $\epsilon_0$ and $n_0$ the initial energy density and number density, respectively. The energy is always conserved during the evolution while the particle number would also be conserved if only elastic scatterings are involved.
The value of the parameter $n \, \epsilon^{-3/4}$ that corresponds to the onset of Bose-Einstein condensation, i.e., to an equilibrium state with vanishing chemical potential,  is obtained by taking for $f(p)$ the  ideal  distribution for massless particles at temperature $T$. One gets then $\epsilon_{SB}=(\pi^2/30)\, T^4$ and $n_{SB}=(\zeta(3)/\pi^2)\, T^3$, so that
\begin{eqnarray}\label{overpopparam}
n \, \epsilon^{-3/4}  |_{SB} = \frac{30^{3/4}\, \zeta(3)}{\pi^{7/2}} \approx 0.28.
\end{eqnarray}
Comparing with $n_0 \, \epsilon_0^{-3/4} $ in Eq.~(\ref{initialparam}), one  sees that when $f_0$ exceeds the value $f_0^c \approx 0.154$, the initial distribution (\ref{eq_f0}) contains too many gluons to be accommodated in an equilibrium Bose-Einstein distribution, i.e. overpopulated: {\it in this case the equilibrium state will have to contain a  Bose-Einstein condensate if there are only elastic scatterings}. It is worth emphasizing that over-occupation does not require necessarily large values of $f_0$, in fact the values just quoted are smaller than unity. It follows therefore that the situation of over-occupation will be met for generic values of $\alpha_s$. For instance, for $\alpha_s\simeq 0.3$, $f_0=1/\alpha_s$ is significantly larger than $f_0^c$ for a wide class of initial conditions (and even more so if the coupling is smaller).  
Such a novel proposal of possible  Bose-Einstein Condensation phenomenon during glasma evolution has generated wide interests and triggered intensive studies about the existence and the  phenomenological roles of such a condensate (see e.g. \cite{Berges:2012ks,Huang:2014iwa} for reviews and references).

While the above thermodynamic consideration implies the connection between the initial state overpopulation and the final state condensate, the questions of precisely how  the thermalization proceeds and how the BEC occurs dynamically, remain interesting and intriguing. The kinetic evolution based upon transport equation  provides a natural and plausible framework for describing the dynamical onset of BEC in the Glasma. In \cite{Blaizot:2013lga}, the kinetic equation with elastic gluon scatterings (which is long-range Coulomb-type interaction)  in the small angle approximation has been derived, which takes the form of a Fokker-Planck equation
describing momentum space diffusion: 
\begin{eqnarray}
\label{bolel}
\partial_t f_1=-{\vec \bigtriangledown}_1\cdot{\bf \cal J}({ \vec p}_1),
\end{eqnarray}
with 
\begin{eqnarray}
\label{current}
{\cal J}( \vec p)=-\frac{g^4N_c^2L}{8\pi^3}\left [ I_a{\vec  \bigtriangledown}f_p+I_b\hat{p} f_p(1+f_p)\right ].
\end{eqnarray}
The two integrals $I_a$ and $I_b$ are defined as follows
\begin{eqnarray}
\label{intia}
I_a\equiv2\pi^2\int \frac{d^3\vec{p}}{(2\pi)^3}f_p(1+f_p) \, , \quad \, 
\label{intib}
I_b\equiv 2\pi^2\int \frac{d^3\vec{p}}{(2\pi)^3}\frac{2f_p}{E_p}.
\end{eqnarray}

\begin{figure}[htbp]
\centering
\vspace{0.3cm}
  \includegraphics[width=5.5cm]{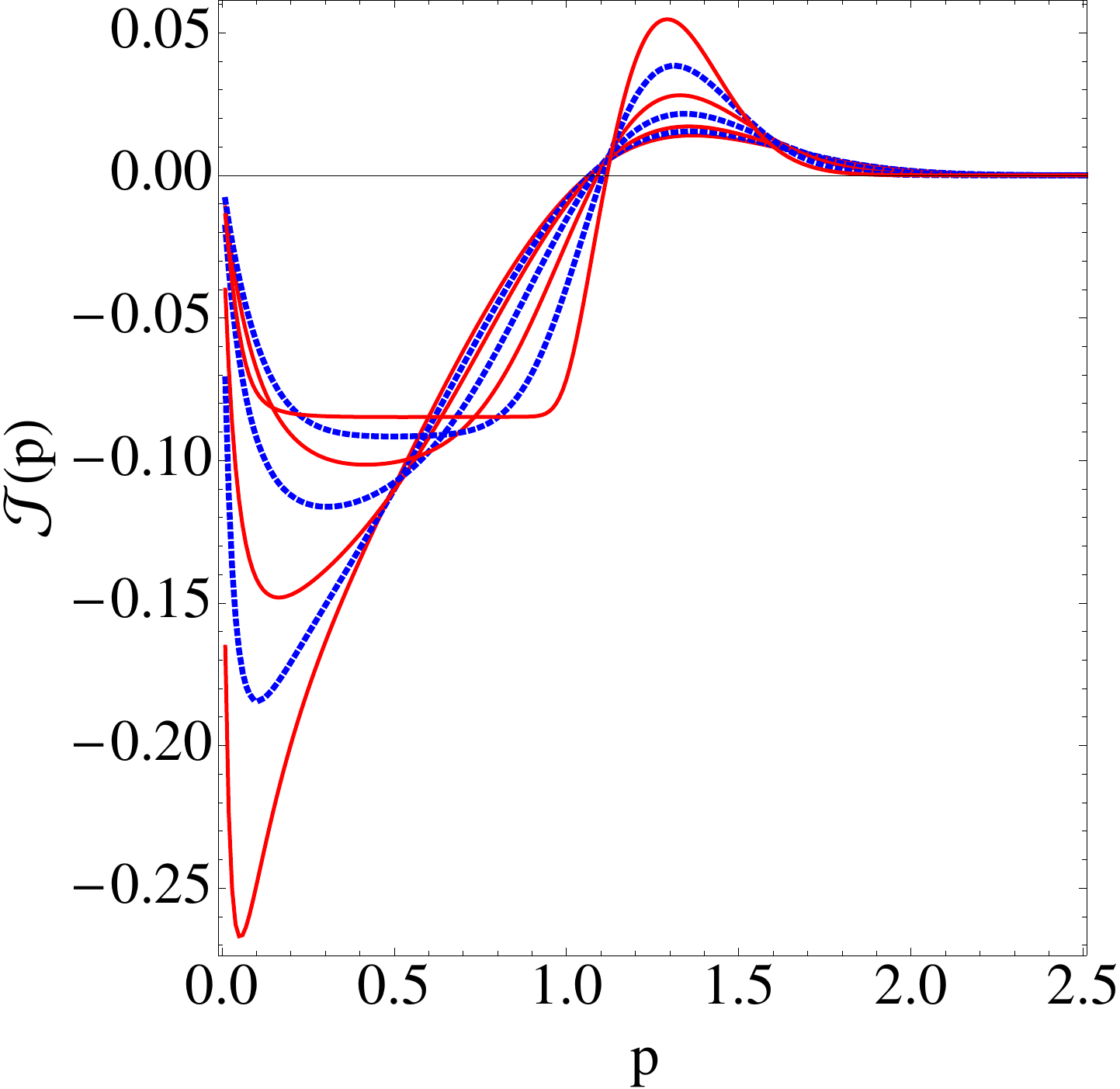} \hspace{0.2cm}
  \includegraphics[width=6.cm]{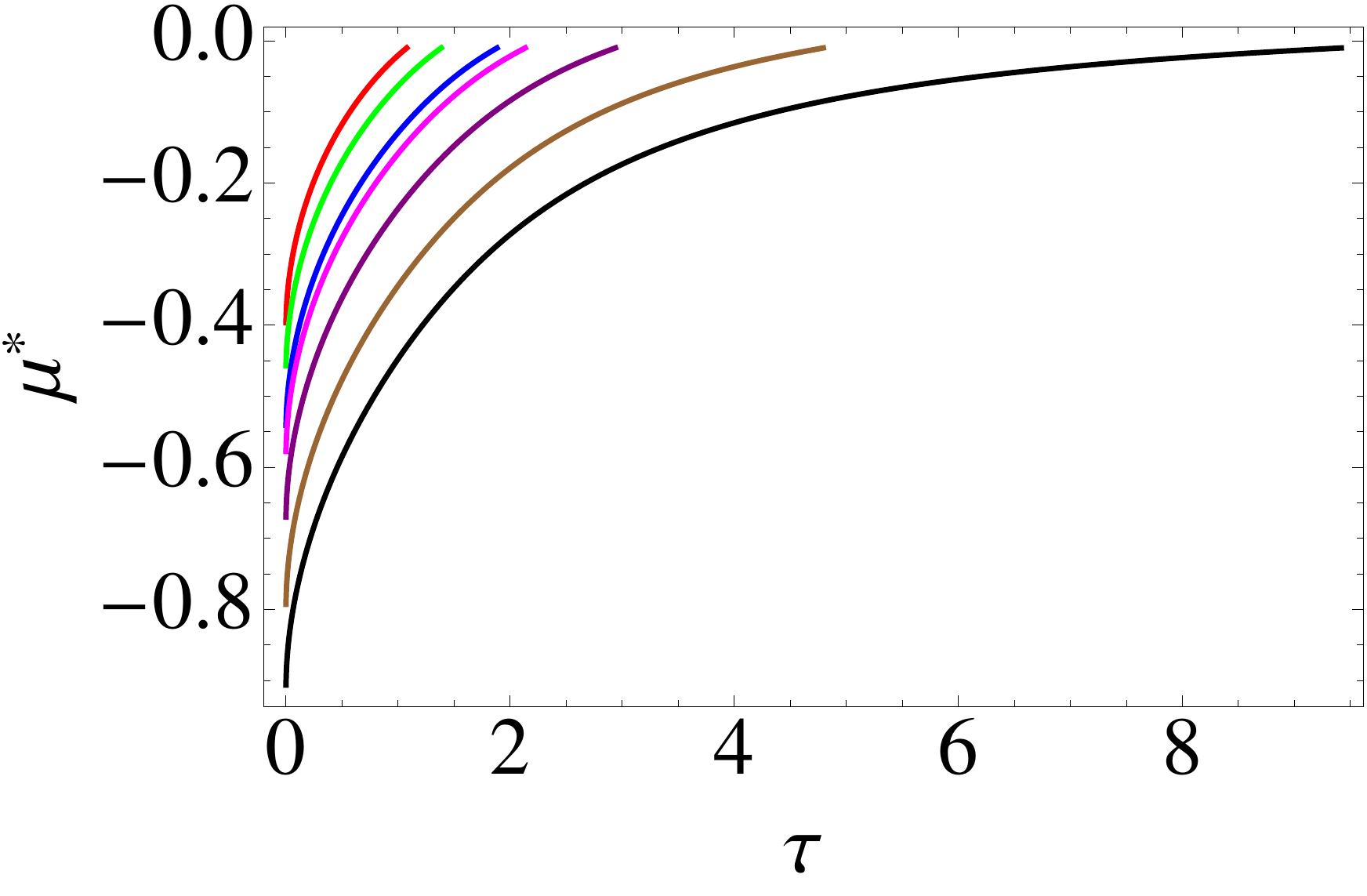}		
		\caption{(left) The the current ${\mathcal J}(p)$ for various times, from an early time till the time near onset (with initial occupation $f_0=1$); (right) The time dependence of local chemical potential $\mu^*(\tau)$  for various initial occupation $f_0$.}
		\label{fig:distr_f1}
\end{figure}

\begin{figure}[htbp]
\centering
\includegraphics[width=10.cm,height=7cm]{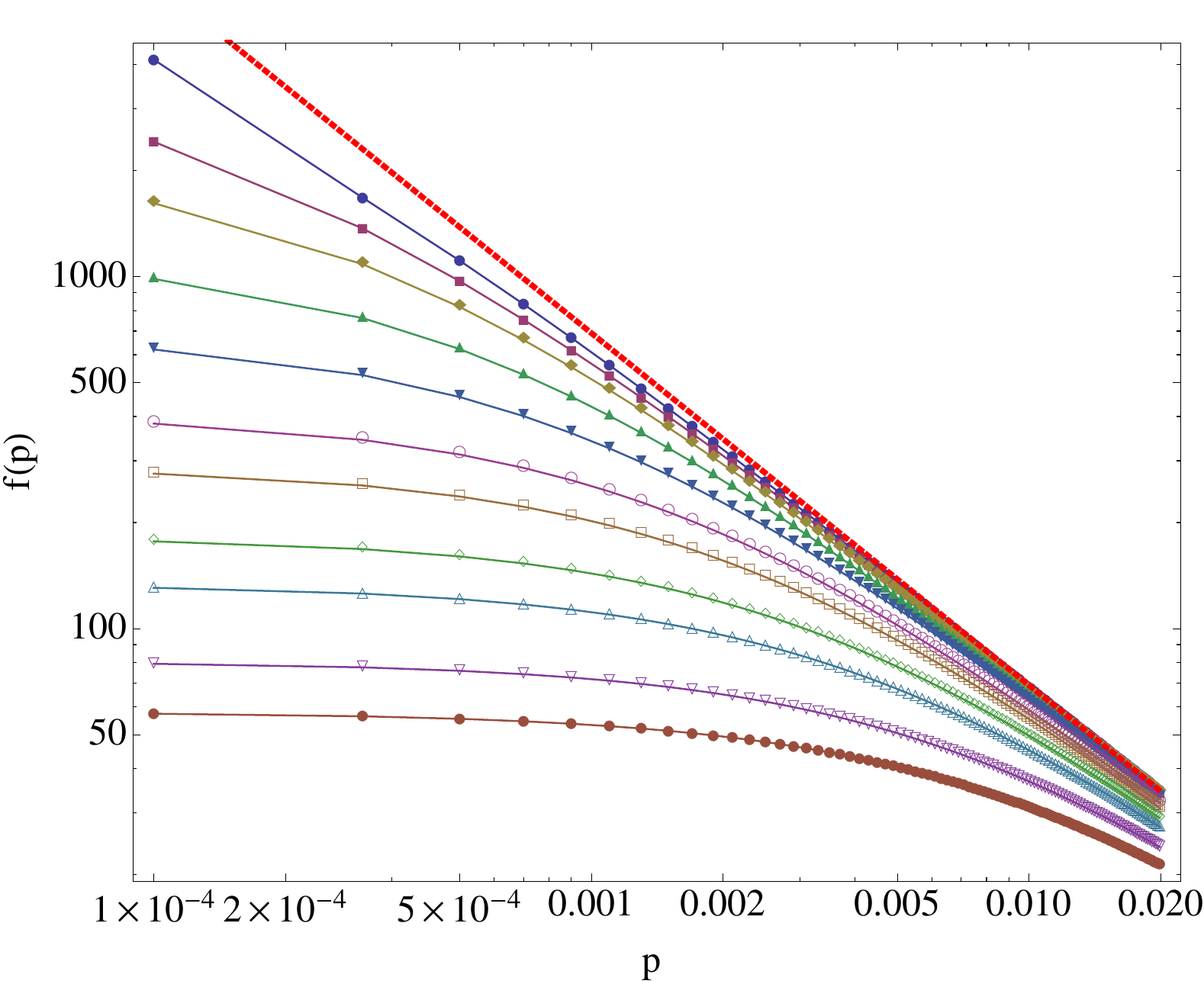}		 	
		\caption{The IR distribution function $f(p)$  for various times, from an early time till the time near onset (with initial occupation $f_0=1$).}
		\label{fig:distr_f1b}
\end{figure}

By numerically solving this equation with varied initial conditions, we have found a few  highly nontrivial and universal features~\cite{Blaizot:2013lga}. First, two cascades in momentum space will quickly develop: a particle cascade toward the IR momentum region that quickly populates the soft momentum modes to high occupation, and a energy cascade toward the UV momentum region that spreads the energy out: see e.g. Fig.\ref{fig:distr_f1} (left panel). The two cascades are of course interrelated as per the particle number and energy conservation. 
Second,  the Bose statistical factors play a key role in the strong particle cascade toward IR, amplifying the rapid  growth of the population of the soft modes. As a consequence a high occupation number at IR is quickly achieved, thus with very fast scattering rate, leads to  an almost instantaneous local IR ``equilibrium'' form for the distribution near the origin $p\to 0$:
\begin{eqnarray}
f^*(p\to 0)=\frac{1}{{\rm e}^{(p-\mu^*)/T^*}-1},
\end{eqnarray}
 The quick emergence of such local IR thermal form has also been numerically verified, see e.g. Fig.\ref{fig:distr_f1b}. Note that the $T^*$ and $\mu^*$ are only parameters characterizing the small momentum shape of the distribution and not to be confused with a true thermal temperature and chemical potential. Finally, starting with overpopulated initial condition (with occupation exceeding a critical value),   the IR cascade persists to drive the local thermal distribution near $p=0$ to increase rapidly in a self-similar form (see Fig.\ref{fig:distr_f1b}). The associated negative local ``chemical potential'' is driven to approach zero, i.e. $(-\mu^*) \to 0^+$ (see Fig.\ref{fig:distr_f1} right panel) and ultimately vanishes in a finite time, marking the onset of the condensation. The approaching toward onset is well described by a scaling behavior:
\begin{eqnarray}
|\mu^*|=C(\tau_c-\tau)^\eta
\end{eqnarray}
with a universal exponent $\eta\approx 1 $. One may analytically show that the exponent is expected to be unity via similar scaling arguments used in the famous turbulent wave scaling analysis.  Such evolution toward onset is robust against different initial distribution shapes, e.g. the same behavior was found with a Guassian initial distribution in the overpopulated regime. 

These results, obtained by using  kinetic theory, with a quantum Boltzmann equation in the small angle approximation, have therefore   provided numerical evidence that  a system of gluons with an initial distribution that mimic that expected in heavy ion collisions reaches the onset of Bose-Einstein condensation in a finite time.  The role of  Bose statistical factors in amplifying the rapid  growth of the population of the soft modes is essential.  With these factors properly taken into account, one finds that elastic scattering alone provides an efficient mechanism for populating soft modes. The general link from initial overpopulation to the onset of BEC in a finite time with a scaling behavior appears to be very robust. Such novel findings are both of their own interests and bearing important implications for the thermalization problem in heavy ion collisions. The kinetic study of dynamical BEC onset in the glasma has recently been extended in a number of aspects: the effects of including inelastic gluon scatterings are studied in \cite{Huang:2013lia}; the inclusion of quarks into the glasma are studied in \cite{Blaizot:2014jna}; the medium effects via dressing the gluons with finite mass are studied in \cite{BJL}. In all cases, the key features of overpopulated glasma, i.e the rapid IR thermalization and the dynamical onset of BEC via vanishing local chemical potential with scaling behavior, are found to be generic and robust.

\subsection{Scaling solution of glasma evolution}

To qualitatively describe the kinetic evolution, one may  introduce two scales for characterizing a general distribution: a soft scale $\Lambda_{\rm s}$ below which the occupation reaches $f (p<\Lambda_{\rm s}) \sim 1/\alpha_{\rm s}\gg 1$ and a hard cutoff scale $\Lambda$ beyond which the occupation is negligible $f (p>\Lambda)\ll 1$. For the glasma initial distribution in (\ref{eq_f0}),  there is essentially only one scale i.e. the saturation scale $Q_{\rm s}$ which divides the phase space into two regions, one with $f\gg1$ and the other with $f\ll 1$, i.e. with the two scales overlapping $\Lambda_{\rm s} \sim \Lambda \sim Q_{\rm s}$. The thermalization is a process of maximizing the entropy (with the given amount of energy). The entropy density for an arbitrary distribution function is given by $s \sim \int d^3\vec{p} \left[(1+f)\, \ln(1+f)-f \, \ln (f)\right]$: this implies that with the total energy constrained, it is much more beneficial to have as wide as possible a phase space region with $f\sim 1$.  Indeed, for a thermal Bose gas one has the  soft scale $\Lambda_{\rm s}^{th} \sim \alpha_{\rm s} T $ and the hard scale $\Lambda^{th} \sim T$ separated by the coupling $\alpha_{\rm s}$. By this general argument, one shall   expect the separation of the two scales along the thermalization process: from the $\Lambda_s \sim \Lambda$ in the initial glasma toward the $\Lambda^{th}_s \sim \alpha_{\rm s} \Lambda^{th}$ in the thermal situation.

To be more quantitative, one may define the two scales $\Lambda$ and $\Lambda_{\rm s}$ as follows:
\begin{eqnarray}
\Lambda \left( {{\Lambda_{\rm s}} \over {\alpha_{\rm s}}} \right)^2  \equiv I_a  \quad &,& \quad
\Lambda \left({{\Lambda_{\rm s}} \over {\alpha_{\rm s}}} \right) \equiv I_b   \\
{\rm or} \;\;\;\;\;\;\;\Lambda = \frac{I_b^2}{I_a} \quad  &,& \quad \Lambda_s = \alpha_s \frac{I_a}{I_b}
\label{eq_def2}
\end{eqnarray}
With the above definition we indeed have $\Lambda_{\rm s} \sim \Lambda \sim Q_{\rm s}$ for the glasma distribution while $\Lambda_{\rm s}^{th} \sim \alpha_{\rm s} \Lambda^{th} \sim  \alpha_{\rm s}  T$ for thermal distribution. Again one  can see  that with the overpopulated  glasma distribution the collision term ${\cal C}\sim \Lambda_s^2 \Lambda \sim \hat{o}(1)$ in coupling, in contrast to the thermal case with ${\cal C}\sim {\Lambda^{th}_s}^2 \Lambda^{th} \sim \hat{o}(\alpha_s^2)$.

Let us now  discuss possible scaling solution for the evolution of the two scales in the static box case. With the glasma distribution the scattering time from the collision integral on the RHS of the transport equation (\ref{bolel}) scales as $t_{\rm sca} \sim \Lambda / \Lambda_{\rm s}^2$. To find scaling solution for the time evolution of $\Lambda$ and $\Lambda_s$, we use two conditions --- that the energy must be conserved and that the scattering time shall scale with the time itself, i.e.:
\begin{eqnarray} \label{eq_sca}
t_{\rm sca} \sim \frac{\Lambda}{ \Lambda_{\rm s}^2} \sim t   \quad  , \quad \epsilon \sim \frac{\Lambda_{\rm s} \Lambda^3 }{ \alpha_{\rm s} } =  {\rm constant}
\end{eqnarray}
The particle number also must be conserved, albeit with a possible component in the condensate:
$n =  n_g + n_c  \sim (\Lambda_{\rm s} \Lambda^2/ \alpha_{\rm s}) + n_c  =  {\rm constant}$.  The condensate plays a vital role with little contribution to energy while unlimited capacity to accommodate excessive gluons. With these two conditions we thus obtain:
\begin{eqnarray}
\Lambda_{\rm s}  \sim Q_{\rm s} \left( \frac{t_0}{t} \right)^{3/7} \quad , \quad  \Lambda  \sim Q_{\rm s} \left( \frac{t_0}{t} \right)^{-1/7}
\end{eqnarray}
From this solution, the gluon density $n_g$ decreases as $\sim (t_0/t)^{1/7}$, and therefore the condensate density is growing with time, $n_c \sim (Q_{\rm s}^3/\alpha_{\rm s}) [1-(t_0/t)^{1/7}]$. A parametric thermalization time could be identified by the required  $\Lambda_{\rm s} / \Lambda \sim \alpha_{\rm s}$:
\begin{eqnarray}
t_{\rm th} \sim \frac{1}{Q_{\rm s}} \,  \left( \frac{1}{\alpha_{\rm s}} \right)^{7/4}
\end{eqnarray}
At the same time scale the overpopulation parameter $n\epsilon^{-3/4}$ indeed also reduces from the initial value of order $\sim 1/\alpha_s^{1/4}$ to be of the order one.

What would change if one considers the case with boost-invariant  longitudinal expansion?
First of all the conservation laws will be manifest differently:  the total number density will decrease as $n\sim n_0 t_0/t$, while the time-dependence of energy density depends upon the momentum space anisotropy $ \epsilon \sim \epsilon_0 (t_0/t)^{1+\delta}$  for a fixed anisotropy
 $\delta \equiv 	P_L/\epsilon$ (with $P_L$ the longitudinal pressure). Along similar line of analysis as before with the new condition of energy evolution  we obtain the following scaling solution in the expanding case:
\begin{eqnarray} \label{eq_scales_expansion}
\Lambda_{\rm s}  \sim Q_{\rm s} \left( t_0/t \right)^{(4+\delta)/7} \, , \,  \Lambda  \sim Q_{\rm s} \left( t_0/t \right)^{(1+2\delta)/7}  \, .
 \end{eqnarray}
With this solution, we see the gluon number density $n_g \sim (Q_{\rm s}^3/\alpha_{\rm s}) (t_0/t)^{(6+5\delta)/7}$, and therefore with any $\delta > 1/5$ the gluon density would drop faster than $\sim t_0/t$ and there will be formation of the condensate, i.e.
$n_c \sim (Q_{\rm s}^3/\alpha_{\rm s}) (t_0/t) [1-(t_0/t)^{(5\delta-1)/7}]$. Similarly a thermalization time scale can be identified through the separation of scales to be:
\begin{eqnarray}
t_{\rm th} \sim \frac{1}{Q_{\rm s}} \,  \left( \frac{1}{\alpha_{\rm s}} \right)^{7/(3-\delta)} \, .
\end{eqnarray}
The possibility of maintaining a fixed anisotropy during the glasma evolution is not obvious but quite plausible due to the large  scattering rate $\sim \Lambda_{\rm s}^2/\Lambda \sim 1/t $ that is capable of competing with the $\sim 1/t$ expansion rate and may reach a dynamical balance. In such a scenario  the system may evolve for a long time with finite anisotropy between average  longitudinal and transverse momenta.

\section{The glow of the glasma}

Phenomenologically, it is of great interest to find any observable that may carry information directly from this pre-thermal glasma stage.  A possible probe is the emission of photons and dileptons which is continuously produced during all stages of heavy ion evolution and once produced, will travel through the fireball largely without interacting with medium due to the small electromagnetic cross sections. Therefore they provide unique access to the history of the quark gluon matter from very early times, including  the pre-thermal glasma evolution.

Measurements of such photons and dileptons produced in collisions at RHIC were first reported by the PHENIX experimental collaboration with quite surprising results \cite{Adare:2008fqa,Adare:2009qk}.  There is a large excess of photons in the transverse momentum range of $1-3~GeV$ for central gold-gold collisions.  This excess far exceeds that due to direct photons and has been interpreted by the PHENIX collaboration to represent photons produced by a Quark-Gluon Plasma with a temperature in excess of the deconfinement temperature.
Recently PHENIX has also measured the flow (or more precisely the azimuthal anisotropy) of such photons  and found it to be sizable and exceeding the expectation from the hydrodynamic expansion of a thermalized Quark-Gluon  Plasma \cite{Adare:2011zr}.  Furthermore an excess of dileptons in the mass range $ 100~ MeV \le M \le 1~GeV $ is also present in the PHENIX data.   In addition, the slope of the $k_T$ distribution for such excessive dileptons in the relevant invariant mass bins is in the range of few hundred MeV,  much less than the corresponding mass of the dileptons: this fact is in sharp contrast with the typical kinematics $k_T \sim M$ if  such pairs were to arise either from a thermal emission from a Quark Gluon Plasma or from semi-hard processes.
 More recently the STAR experimental collaboration has also measured the low to intermediate mass dileptons~\cite{wang,zhao} and the reported results show much less excess than that reported by PHENIX.  At the moment it is not clear yet whether the STAR measured enhancement  of these pairs  have their origin in low transverse momentum region. Many phenomenological studies (e.g. \cite{rapp_RHIC,vHvRapp,dusling_RHIC,cassing_RHIC,Shen:2013vja}) have been done in order to understand the implications of these data.

As the experimental data hint at possible contribution from unconventional source, it is of great interest to examine the electromagnetic emissions from the pre-thermal glasma stage, which has not been much studied. A first step was made in \cite{Chiu:2012ij} to roughly estimate  the production of photons and dileptons during the pre-equilibrium glasma stage in heavy ion collisions, by applying the scaling description of glasma evolution discussed in the previous Section.

\subsection{Photons} 
 
Let us first discuss the photon production, through the Compton channel $qg\to q\gamma$, with the quark and gluon distributions characterized by the two scales $\Lambda_s$ and $\Lambda$ in the glasma as previously introduced. The glasma emission rate is then parametrically given by:   
\begin{equation}
  {{dN} \over {d^4x dy d^2k_T}} = {\alpha \over \pi} \Lambda_s \Lambda \, g(E/\Lambda) \label{photratefb}
\end{equation}
Here, $g$ is a function of order one that cuts off when then energy of the photon is of the order
of the UV cutoff scale $\Lambda$. Note that the usual $\alpha_s$ factor from quark-gluon coupling is compensated by the $1/\alpha_s$ factor from the gluon occupation. In obtaining the above, we have assumed that the quarks will be generated (e.g. via gluon annihilations ) on a time scale of $1/Q_s$ and following a distribution of the form $F_q(p/\Lambda)$ (see discussions in~\cite{Chiu:2012ij}) that is an order one function characterized by the UV scale. These are plausible assumptions provided our current understanding of quark generation in the glasma, see e.g. dedicated studies of  this problem in \cite{Blaizot:2014jna,Gelis:2013oca}. One then needs to integrate out the space-time volume. To do the time integral one uses the time-dependence of the two scales in (\ref{eq_scales_expansion}). At the end one obtains the parametric photon yield to be
\begin{equation}
{dN_{\gamma} \over dy\,d^{2}k_{T}} \sim \alpha R_{0}^{2}N_{part}^{2/3}\left ( {Q_{sat} \over k_{T}} \right )^{\eta}\,,
\label{eqn_phot1}
\end{equation}
where \,$\eta = (9 - 3\delta)/(1 + 2\delta)$ is the parameter from (\ref{eq_scales_expansion}) related to the anisotropy of glasma expansion. $R_0$ is a constant with the dimension of length (introduced upon scaling the overlap zone area with $N_{part}^{2/3}$). Finally, by properly implementing scaling properties of the involved scales, one arrives at the following phenomenological formula for photon yield from the glasma: 
\begin{eqnarray} \label{eq_photon}
{dN_{\gamma} \over dy\,d^{2}k_{T}}  = C_\gamma\, N_{part}^{2/3} \times {\left [ Q_{0}^{2}\left( \sqrt{s}\times 10^{-3} \right)^{\lambda} \right ]^{\eta/2}
\over k_{T}^{\,\eta\,(1+\lambda/2) }}  \,   .
\end{eqnarray} 
with $\lambda$ a parameter charactering the evolution of scales with $x$. The  constant coefficient $C_\gamma \propto \alpha R_0^2$ can be determined by fitting at one centrality bin and then applied to all other centralities.  A detailed discussion of all these formulae can be found in \cite{Chiu:2012ij}.

\begin{figure}[htbp]
	\begin{center}
		\includegraphics[width=8.cm]{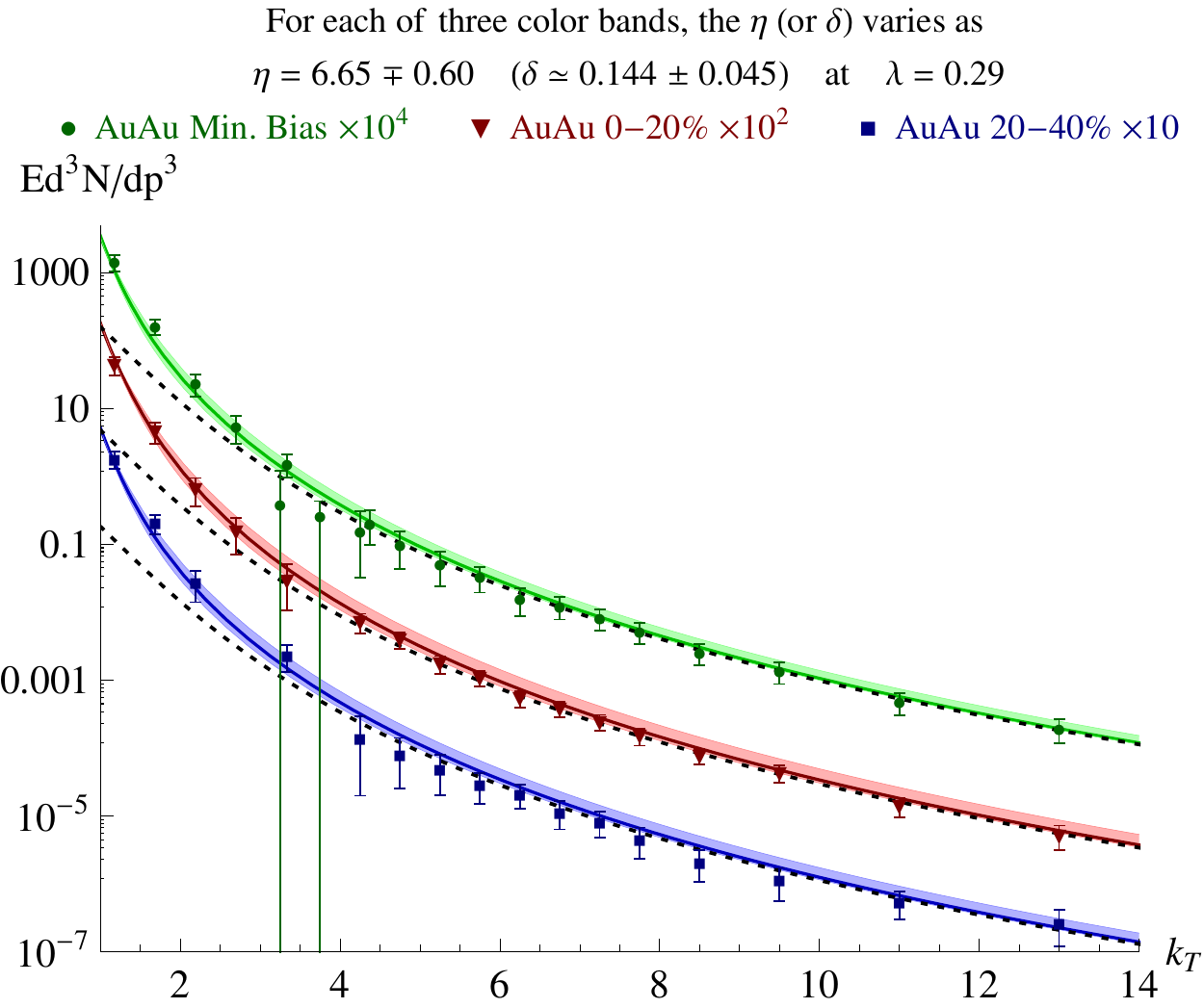}  
		\caption{  The $k_T$ spectra of photon production including emissions from the pre-equilibrium matter on top of convention sources. The dashed lines are contributions from conventional sources, while the color bands represent the total production including pre-equilibrium emissions, and the data points are from PHINIX measurements (see \cite{Chiu:2012ij} for details). }	\label{fig_photon}
	\end{center}
\end{figure}

We then add this glasma contribution on top of conventional yield formula parameterized by Hagedorn function in pp collision properly scaled to AA collision.  
The data we aim to describe will be the invariant yield in $Au$-$Au$ collisions
at \,$\sqrt{S_{NN}} = 200\,GeV$\, of direct photons at centralities \,$0$-$20^{\circ\!\!}/_{\!\!\circ}$,\,
\,$20$-$40^{\circ\!\!}/_{\!\!\circ}$\, and \,$0$-$92.2^{\circ\!\!}/_{\!\!\circ}$ (Min. Bias) as a function
of $k_{T}$ determined  from the PHENIX measurement: see Fig.\,$34$ of Ref.\,\cite{Adare:2009qk}. The strategy is the following: for given values
of $\lambda$ and $\eta$ in Eq.(\ref{eq_photon}), we will fix the coefficient $C_\gamma$ from the $0$-$20^{\circ\!\!}/_{\!\!\circ}$ case and test how well the formula describe the data at the other two centrality choices. This will provide a critical test of the geometric scaling properties of the present model. The best description of data is obtained with $\lambda=0.29$ and $\eta=6.65$ (corresponding to $\delta=0.144$). The parameter $C_\gamma$ is fitted to be $ 0.0234 fm^2$. The results are presented in Fig.\ref{fig_photon}. As one can see, the inclusion of the glasma contribution is necessary and even dominant for describing the ``excess'' in yield and the $k_T$-dependence in the  region about $1\sim 3\, \rm GeV$. Furthermore, the glasma hypothesis yields a simple and robust estimate for the dependence of photon production on centrality as a consequence
 of geometric scaling of the emission amplitudes, which agrees well with  data. Most recently there have appeared further studies of such geometric scaling of photon production based on initial saturation scale~\cite{Klein-Bosing:2014uaa,McLerran:2014hza}.

\subsection{Dileptons}

We now turn to the dilepton emission. The usual process would be the annihilation of quark anti-quark pairs. However as identified in \cite{Chiu:2012ij},  for the interesting ``excess'' seen  in the range between a few hundred $\rm MeV$ and about $1~\rm GeV$,  there is another more important contribution in the glasma with a condensate.  This process is the annihilation of gluons into a quark loop from which the quarks then subsequently decay into a virtual photon and eventually the dilepton.  Such a virtual process is naively suppressed by factors
of $\alpha_s$.  Here however, the gluons arise from a highly coherent condensate, and the corresponding
factors of $\alpha_s$ are compensated by inverse factors $1/\alpha_s$ from the coherence of the condensate.   In other words, the usual power counting for diagrams in terms of $\alpha_s$ has to be changed when the coherent condensate with high occupation is present.
This annihilation process from a condensate has a distinctive feature.  The condensate gluons have nearly zero total momentum in a co-moving frame.  All of the gluon momentum is acquired by collective flow, and hence
the produced dileptons will have a small transverse momentum  which could be much smaller compared with the pair mass. The dilepton differential yield in both invariant mass $m_{ee} \equiv M$ and transverse momentum $k_T$ from this contribution can be estimated by the following phenomenological formula: 
\begin{equation}
{dN_{C \rightarrow DY} \over d^{2}k_{T}dy\,dM^{2}} = C_{ll}
N_{part}^{2/3}\left ( {Q_{sat} \over M} \right)^{\eta^{\prime}}{e^{-k_{T}/\mu} \over \mu^{2}}\,,
\label{eqn_dilep1}
\end{equation}
where the exponent now becomes $\eta'$ given by
$     \eta^\prime = 9(3-\delta)/(5+3\delta)$. As in the case for photon, the scaling of $Q_s$ is given by 
$
Q_{sat}^{2}(k_{T}/\sqrt{s}) =  Q_{0}^{2}\left( \sqrt{s}\times 10^{-3} \over {k_{T}}
\right)^{\lambda}e^{\lambda y}\,  
$, 
which includes its evolution with the transverse momentum $k_{T}$ as well as the rapidity $y$. In the above we have also introduced an exponential function for $k_T$-dependence with
a width scale parameter $\mu$, which upon full integration over $k_T$ will go back to the expected integrated yield (see \cite{Chiu:2012ij}). 
While ideally the dilepton production from condensate would generate a delta
function at $k_T = 0$, there however would be broadening in the pair $k_T$ due to both the finite
transverse size of the system and the transverse collective expansion, which is accounted for by
the introduced exponential function. The width scale parameter $\mu$ is one of the few fitting parameters to be optimized later. Again a detailed discussion of all these formulae can be found in \cite{Chiu:2012ij}.

\begin{figure}[htbp]
	\begin{center}
				\includegraphics[width=8.cm]{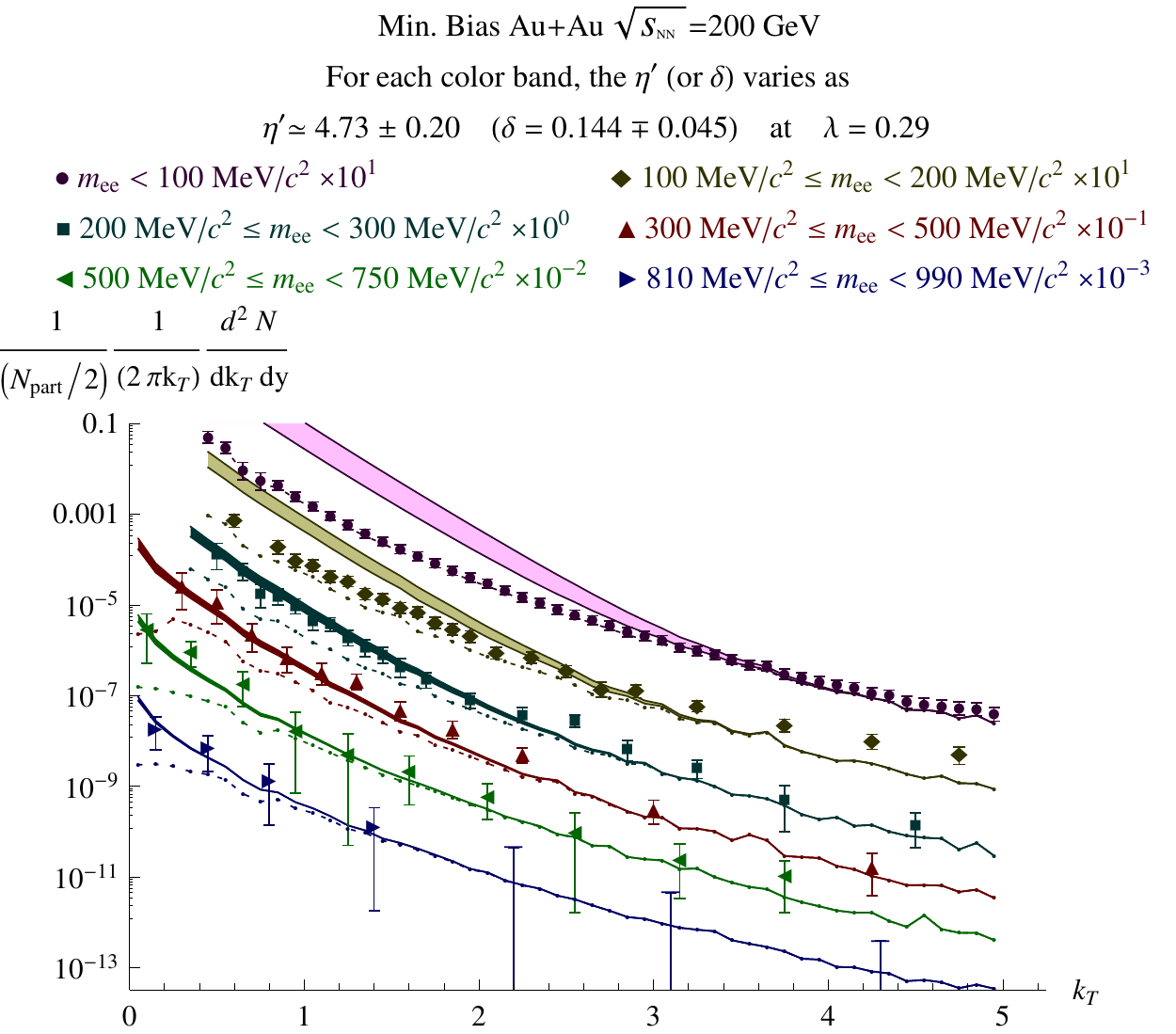}
		\caption{  The $k_T$ spectra of dilepton  production including emissions from the pre-equilibrium matter on top of convention sources. The dashed lines are contributions from conventional sources, while the color bands represent the total production including pre-equilibrium emissions, and the data points are from PHINIX measurements (see \cite{Chiu:2012ij} for details). }	\label{fig_dilepton}
	\end{center}
\end{figure}

The data we aim to describe are for $k_{T}$-dependence of  $e^{+}e^{-}$ pairs at various given mass bins, as measured by PHENIX in $Au+Au$ \,$200\,GeV$ minimum bias collisions.  For a given mass bin $[M_{min}\,,M_{max}]$, we can integrate the differential yield to obtain the $k_T$-spectra. It shall be emphasized that in our model the dileptons generated from the Glasma shall have their mass bounded by the in-medium mass of gluons in the Glasma, and therefore our formula shall not be applied to too small values for the pair mass. Accordingly, we focus on comparison with data in a mass regime $0.2 ~ GeV \le M \le 1~GeV$. For the parameters we will use the same values for $\delta=0.144$ (thus $\eta'=4.73$) and $\lambda=0.29$ determined in the photon production. For the $k_T$ width we have found the optimal value $\mu=0.2\,GeV$. The fitting quality has a relatively strong dependence on $\mu$. The optimal value for the overall normalization $C_{ll}$ determined from such fitting is $C_{ll}\approx 3.5 \times 10^{-6}\,GeV^{-2}$. This coefficient $C_{ll}$ is plausibly expected to be parametrically much smaller than the $C_{\gamma}$ in the photon case, as the former has one more power in its dependence on the electromagnetic coupling $\alpha=1/137$. Our fitting results for the PHENIX dilepton data are shown in Fig.\,\ref{fig_dilepton}. For each mass bin of the dilepton yield, the dashed curves represent the background yield formed by contributions of the hadronic decay cocktail and charmed mesons. For the four higher mass bins relevant to the production from the glamsa, we add on top of the background the additional contribution computed from Eq.\,(\ref{eqn_dilep1}): the results for the total yield are represented by the colorful bands where the upper and lower boundary curves for each band correspond to the results with the parameter $\eta'=4.93$ and $\eta'=4.53$, respectively. The mass-differential spectra were also studied and compared with data and other models: see \cite{Chiu:2012ij} for details. The results of this phenomenological study suggest that:  (1)  The transverse momentum and mass spectra  can be described within the glasma hypothesis;  (2) The observation by PHENIX that the low to intermediate mass enhancement arises from anomalously small transverse
 momentum region has a natural interpretation as emissions from a condensate in the glasma which  in a fluid at rest produces particles with zero transverse momentum. Based on our model, the geometric scaling is predicted for the centrality dependence of the dilepton spectra which can be tested with future data.

\section{Summary}

In summary, we have discussed the evolution of the pre-thermal gluonic matter, the glasma, at early time in heavy ion collisions and the electromagnetic emissions during this stage. A detailed understanding of the approach to equilibrium of the glasma and its phenomenological consequences, remains a big challenge and an active, rapidly developing direction. 

It has been suggested recently that the initial high occupation $f\sim 1/\alpha_s$ of the gluons plays a key role,  making the glasma a weakly coupled but strongly interacting system. In particular such an overpopulated glasma is predicted to develop a dynamical Bose-Einstein Condensate based on rather general argument. A kinetic framework is developed to study the detailed dynamics of such onset of BEC starting with overpopulated initial conditions, which finds several generic features:  strong IR particle cascade and UV energy cascade; rapid emergence of IR local thermal form; dynamical BEC onset via vanishing of local chemical potential in a finite time with a scaling behavior. Studies including inelastic processes, quark contributions, medium screening mass, as well as expansion, have all found these features to be robust.

One possible way to experimentally access the early time system is via the photons and dileptons. It is found that the measured distributions of such electromagnetic emissions, while having some features not well understood if hypothesized to entirely arise from a thermalized Quark-Gluon Plasma, have some qualitative  features that might  be described after including effects from a thermalizing glasma. 
The shape and centrality dependence of the transverse momentum spectra  of the
 so-called "thermal photons" are well described. The mass and transverse momentum dependence of intermediate mass dileptons also agree with our estimates. 
 The low transverse momenta from which the excessive  dileptons (in low to intermediate mass region) arise is suggestive of emissions from a Bose condensate.
 The centrality dependence of dilepton production is also predicted. Clearly the phenomenological study here with rough estimates  is only a first step in the attempts to understand the glasma contributions to electromagnetic emissions. One should bear in mind caveats such as the crudeness of approximations made, the possible alternative explanations as well as the current uncertainty in experimental data (e.g. the low-mass dileptons). With more future progress to provide a detailed understanding of the thermalizing glasma, we expect future studies to make a more quantitative determination of its electromagnetic emissions and to draw a conclusion on its contribution in the observed extraordinary glow in heavy ion collisions.

\section*{Acknowledgements}

The author thanks all his collaborators on the works reported here, in particular J.-P. Blaizot and L. McLerran. He is also grateful to the organizers of this wonderful conference in memory of Gerald E. Brown.  The author's research is supported by the National Science Foundation under Grant No. PHY-1352368. He is also grateful to the RIKEN BNL Research Center for partial support.


\end{document}